\documentclass[journal]{IEEEtran}
\ifCLASSINFOpdf
\else
\fi
\usepackage{graphicx}
\usepackage{subcaption}
\usepackage{color,soul}

\begin{document}

%
\title{Exposing the Robustness and Vulnerability of Hybrid 8T-6T SRAM Memory Architectures to Adversarial Attacks in Deep Neural Networks}
%
%
%

\author{Abhishek~Moitra, and
        Priyadarshini~Panda,~\IEEEmembership{Member,~IEEE}
\thanks{Abhishek Moitra and Priyadarshini Panda are with the Department
of Electrical Engineering, Yale University.}}

%
%

\maketitle
%
\begin{abstract}
Deep Learning is able to solve a plethora of once impossible problems. However, they are vulnerable to input adversarial attacks preventing them from being autonomously deployed in critical applications. Several algorithm-centered works have discussed methods to cause adversarial attacks and improve adversarial robustness of a Deep Neural Network (DNN). In this work, we elicit the advantages and vulnerabilities of hybrid 6T-8T memories to improve the adversarial robustness and cause adversarial attacks on DNNs. We show that bit-error noise in hybrid memories due to erroneous 6T-SRAM cells have deterministic behaviour based on the hybrid memory configurations ($V_{DD}$, 8T-6T ratio). This controlled noise (surgical noise) can be strategically introduced into specific DNN layers to improve the adversarial accuracy of DNNs. At the same time, surgical noise can be carefully injected into the DNN parameters stored in hybrid memory to cause adversarial attacks. To improve the adversarial robustness of DNNs using surgical noise, we propose a methodology to select appropriate DNN layers and their corresponding hybrid memory configurations to introduce the required surgical noise. Using this, we achieve 2-8\% higher adversarial accuracy without re-training against white-box attacks like FGSM, than the baseline models (with no surgical noise introduced). To demonstrate adversarial attacks using surgical noise, we design a novel, white-box attack on DNN parameters stored in hybrid memory banks that causes the DNN inference accuracy to drop by more than 60\% with over 90\% confidence value. We support our claims with experiments, performed using benchmark datasets-CIFAR10 and CIFAR100 on VGG19 and ResNet18 networks.
\end{abstract}


%
\IEEEpeerreviewmaketitle

\section{Introduction}
\IEEEPARstart{D}{eep} Learning has conquered a significant portion of modern day artificial intelligence applications. However, recent works show that they can be easily fooled by making very minimal changes on the input data \cite{szegedy2013intriguing}. What's far worse is that, these minimal changes are invisible to the naked eye. The result is that, the network makes a wrong prediction with a very high confidence. Such failure can prove to be catastrophic for critical applications like medical diagnostics, autonomous driving and stock markets. Hence, the deployment of Deep Neural Networks (DNNs) in real-time applications can be hindered due to the security risks of adversarial attacks.

Since the advent of this problem, there have been extensive works in literature focused on improving the adversarial robustness of DNNs. One of the most promising solution is adversarial training which achieves state-of-the-art results \cite{goodfellow2014explaining, madry2017towards}. Here, the network is trained on adversarial inputs in order to improve the model's robustness. Other works have tried to mitigate the effects of adversarial attacks through image compression \cite{dziugaite2016study}, random input resizing and randomization \cite{xie2017adversarial, xie2017mitigating}, and random gaussian noise augmentation during training \cite{he2019parametric}. Naveed et al. and Yuan et al. describe many other recent works in this direction \cite{MIT_survey, yuan2019adversarial}. We would like to note that most prior works are algorithm-centered and improve adversarial performance by focusing on better training or data optimization strategies.

Efficient hardware implementation of DNNs have shown promising results in terms of improved speed and energy efficiency \cite{finn,xcel_ram,resparc}. For instance, approximate computing leverages the fact that DNNs have a high tolerance towards inaccurate calculations. Thus, we can approximate underlying DNN computing blocks, such as, memory or Multiply-and-ACcumulate (MAC) circuits to minimise energy consumption without affecting the application-level accuracy \cite{mehrotra1994fault, moons2016energy, chakraborty2020constructing, panda2020invited, venkataramani2015approximate}. So far, most hardware related research on DNNs have been focused on trading off energy-and-accuracy. 

There are however, two interesting directions that still remain under-explored i) Using hardware approaches to improve the adversarial robustness of DNNs ii) Exposing the potential vulnerabilities of these hardware approaches specifically towards adversarial attacks. Works such as \cite{bhattacharjee2020rethinking} show that hardware-based approaches can improve the adversarial robustness of DNNs. Other recent works have also shown that homogeneous and heterogeneous (or layer-specific) quantization, can also improve the adversarial accuracy of a DNN \cite{quanos, panda2019discretization}. These works inform that hardware optimization strategies, used for reducing energy consumption, can be used effectively to address the software vulnerabilities in DNNs, specifically adversarial attacks. Other works have shown that hardware attacks can be made on DNN accelerators to cause them to malfunction resulting in serious performance degradation \cite{hw_trojans, memory_attacks1, iia}. These works exploit the structural vulnerabilities of the hardware accelerator such as the micro-architecture and memory access patterns. However, none of these works have employed hardware-based adversarial attacks.

Based on the above insight, in our paper, we take an approximate computing route and explore the robustness and vulnerabilities of hybrid memory architectures towards adversarial attacks on DNNs. Hybrid memories are a combination of 8T and 6T SRAM cells with the most significant bits (MSBs) stored in the 8T cells and the least significant bits (LSBs) stored in the 6T cells. They have been shown to be highly energy efficient for DNN acceleration with minimum performance degradation. This is attributed to its operation under reduced (or scaled) supply voltage. However, at scaled voltages, the 6T SRAM cells behave erroneously leading to bit-error noise which is considered a major drawback of hybrid memories. In this work however, we ask, \textit{``Can the bit-error noise due to LSB 6T-SRAM cells, be introduced in a controlled manner into specific sections of the hybrid memory to improve the adversarial robustness of DNNs? Also, can these noises be used to adversarially attack hybrid memory architectures?"}

We find that for different configurations of 8T-6T ratios and scaled $V_{DD}$, the bit-error noise introduced in the hybrid memory is bound within specific limits. In this paper, we call this strategically crafted bit-error noise as the \textbf{surgical noise}. We show that the surgical noise can be deliberately added in a controlled way into specific sections of the hybrid memory through the process of \textbf{surgical noise injection (SNI)} that can potentially interfere with the creation of adversarial attacks in DNNs, yielding robustness. It is evident that our method preserves the energy-efficiency benefits of hybrid memories due to low $V_{DD}$ operation. Furthermore, we do not require any re-training or training with adversarial data \cite{dq,quanos,vortex}. 

At the same time, we show that SNI can be used to generate adversarial attacks on hybrid memory architectures. To demonstrate this we design a novel white-box surgical noise-based DNN parameter attack, where we assume that the attacker has complete knowledge of the network architecture and parameters of the model. With SNI on specific sections of the hybrid memory storing the DNN parameters, perturbations in the parameter values cause significant performance degradation. Interestingly, we find that these perturbations are adversarial in nature causing the DNN model to incur high confidence mis-classifications.

In summary the novel contributions of our work are as follows:
\begin{enumerate}
    \item We empirically show that the bit-error noise due to LSB 6T-SRAM cells at scaled voltages $V_{DD}$ is a function of 8T-6T cell ratio $r$ and $V_{DD}$. We refer this strategically crafted bit-error noise as surgical noise.
    \item We show that surgical noise in hybrid memory architectures can improve the adversarial robustness of the DNN model when compared to baseline models- without any surgical noise. 
    \item We also bring out a potential vulnerability of hybrid memories by devising a novel white-box, surgical noise-based DNN parameter attack wherein, surgical noise is introduced into specific sections of the hybrid memory storing the DNN parameters which cause high confidence mis-classifications by the model.   
\end{enumerate}

To validate our proposed methodology to improve adversarial robustness using hybrid memories, we run experiments using CIFAR10 and CIFAR100 datasets on both VGG19 and ResNet18 network architectures. Likewise, we validate our proposed adversarial, surgical noise-based DNN parameter attack scheme using the CIFAR10 dataset on VGG19 and ResNet18 architectures.

\section{Related Works}
\label{Related works}
Since the advent of adversarial attacks, there have been numerous algorithm-centric works in the direction to abate such attacks. The most prominent work relates to adversarial training \cite{domain_adv_training,tramer2017ensemble}. This technique of introducing adversarial examples during training, by far has shown very promising results. Specifically, adversarial training with Projected Gradient Decent (PGD) attacked examples has been shown to be the most effective \cite{athalye2018obfuscated,cisse2017parseval}. Other works involve input gradient regularization to minimize the effect of adversarial inputs on the DNN \cite{ross2017improving} and input denoising \cite{xie2019feature} that designed new network architectures that minimize the noise on the input examples for better robustness.

Recently hardware focused approaches like the one by Panda et al. showed that discretization of input examples improves the adversarial robustness of a DNN manifolds. Also, binary neural networks are shown to be more adversarially robust than their full precision counterparts \cite{panda2019discretization}. Works such as QUANOS and Defensive Quantization (DQ) train the network to obtain the best optimized data quantization for improved adversarial robustness \cite{quanos,dq}. While DQ uses homogeneous data quantization for all the layers of the DNN, QUANOS uses adversarial noise sensitivity to determine layer specific data quantization values. The work by Rakin et al \cite{he2019parametric} uses gaussian noise injection on the DNN parameters during the training phase to improve the adversarial robustness of the trained model. The authors show that adding the DNN parameter noise acts as a regularizer. This helps the network to generalize well over the dataset and achieve a higher adversarial accuracy when combined with PGD adversarial training.

Hybrid memories have been extensively used for energy efficient and approximate storage. Works by Chang et al and Bortolotti et al have used hybrid memories to minimize the energy consumption of tasks like MPEG video processing and bio-signal processing respectively \cite{chang2011priority,bortolotti2014approximate}. In the light of approximate computing for deep learning applications, Srinivasan et al showed that hybrid memories can be used for energy efficient computations while maintaining good performance in DNN accelerators \cite{hybridmem}. The authors also show that using different 8T-6T ratios for different layers in the DNN can yield better energy efficiency at iso-accuracy. 

Over time, research in the field of hardware-based side-channel attacks have shown that hardware trojans \cite{hw_trojans,liu2017neural} can degrade the performance of DNN hardware accelerators while remaining extremely stealthy and difficult to detect. In the works by Tolulope et al and Zhao et al, hardware trojans have been used to attack DNN accelerators by analysing the memory data patterns \cite{memory_attacks1,iia}. Recently, a work by Kim et al has shown that DNN performance is adversely affected by frequent accesses of the DRAM memory storing the DNN parameters, which cause bit-flipping \cite{kim2014flipping}. They call it the Row-Hammer attack. Further, Rakin et al \cite{rakin2019bit} designed a progressive bit-search algorithm to detect the most sensitive bits for such bit-flip based attacks.     

In this work, we use bit-error noise in hybrid memories to improve the adversarial robustness of the model which is to the best of our knowledge, the first of its kind in the literature. On top of that, we also show that hybrid memories are vulnerable to adversarial bit-error noise attacks which we demonstrate by designing a novel, white-box attack on the DNN parameters stored in the hybrid memory.

\section{Background}
\label{backgroud}
\subsection{Adversarial Attacks}
\label{adversarial_attacks}
In recent times, various adversarial attack methods have been developed to completely fool a DNN by introducing visibly minor perturbations in the input image data. Such adversarial attacks can be classified into two categories: white-box attacks and black-box attacks.

White-box attacks are based on attacking the target model with complete knowledge of the model architecture and parameters. On the other hand, black-box attacks are realised when the attacker has no knowledge of the target model's network architecture and parameters. It must be noted that robustness against white-box attacks ensures robustness under black-box attacks with similar strengths ($\epsilon$). In this paper, we refer to \textit{clean accuracy} as the accuracy when the DNN model is fed clean inputs without any adversarial perturbations. Similary, \textit{adversarial accuracy} refers to the model accuracy when the DNN is fed with images perturbed with adversarial disturbances. Our objective is to improve the \textit{adversarial accuracy}, which determines the adversarial robustness of the model.
\subsubsection{Fast Gradient Sign Method (FGSM)}
FGSM attack is one of the most efficient and powerful single-step attack. The generation of the adversarial example $\hat{x}$ from clean example $x$ with target value $t$, and DNN parameters $\theta$ follows the equation: $$\hat{x} = x + \epsilon . sign(\nabla_{x} \mathcal{L}(g(x_{i};\theta),t)) $$ The value of $\epsilon$ determines the strength of the adversarial attack. Note that the perturbations are created in the direction of the gradient $\nabla_{x} \mathcal{L}(g(x_{i};\theta),t)$ such that it increases the loss, thereby causing adversarial effects. It has been shown that even with small perturbation values, the DNN model incurs high confidence mis-classifications. 
\subsection{Hybrid memories}
\begin{figure}[h!]
\includegraphics[width=0.5\textwidth]{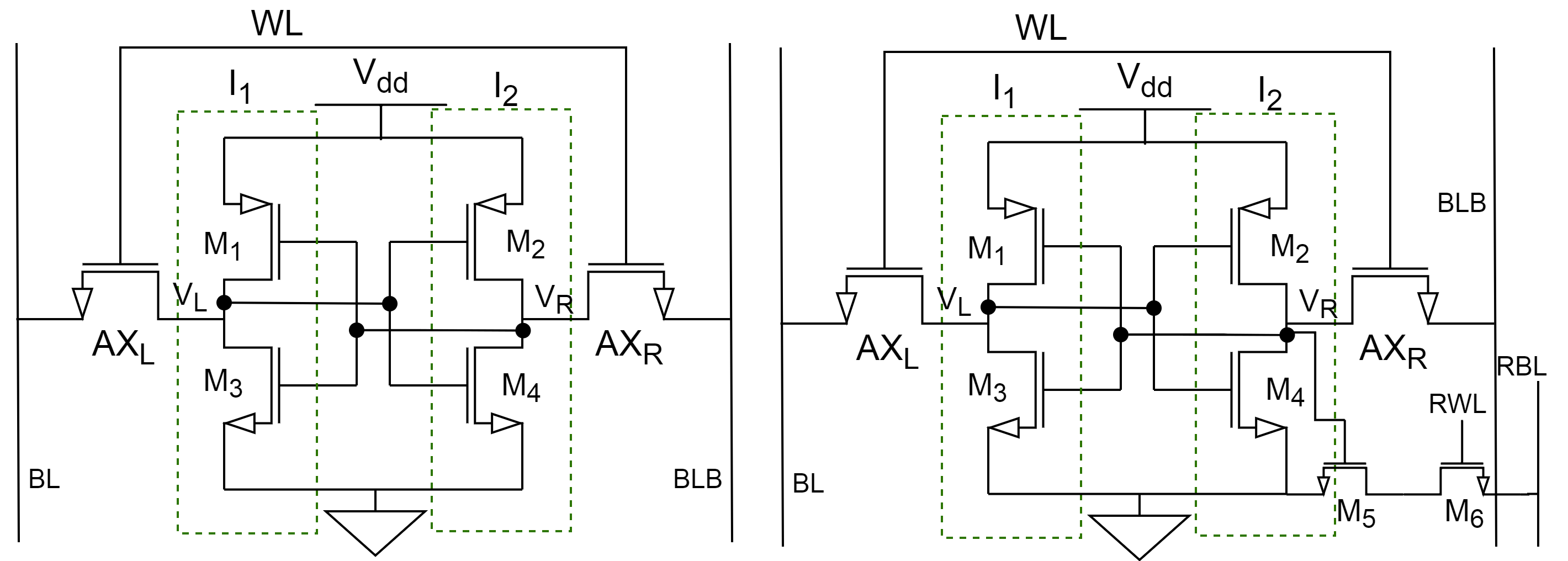}
\caption{Structure of (left) 6T-SRAM having six transistors and common read write bit-lines (BL/BLB) and (right) 8T-SRAM cell having eight transistors and separate read write bit-lines (BL/BLB and RBL)}
\label{sram}
\end{figure}
The most ubiquitous SRAM memory architecture is one having a homogeneous 6T SRAM shown in Fig. \ref{sram}(left). The reason for its popularity is the small silicon footprint which implies higher data storage. However, they suffer from read/write bit-errors at scaled voltages ($V_{DD}$) which is because of the single read/write line ($BL/BLB$). On the other hand, the 8T SRAM shown in Fig. \ref{sram}(right) is nearly unaffected by voltage scaling which makes it suitable for critical applications like aerospace where error resilience is of prime importance \cite{chang20088t}. 

Hybrid memories concoct the area efficiency of 6T-SRAM cells and resilience of 8T-SRAM cells at scaled voltages to yield an energy efficient approximate computing solution \cite{chang2011priority, bortolotti2014approximate}. With regard to DNNs, it has been shown that hybrid memories facilitate aggressive voltage scaling while retaining high accuracy performance which can be used to reduce the energy consumption in DNN accelerators \cite{hybridmem}. Throughout this paper, we refer to the 8T-6T ratio $r$ and $V_{DD}$ as the hybrid memory configurations. The hybrid memory configurations ($r$, $V_{DD}$), can vary or remain constant across different DNN layers.
\subsection{DNN Accelerator Architecture}
\label{dnn accelerator}
\begin{figure}[h!]
\includegraphics[width=0.5\textwidth]{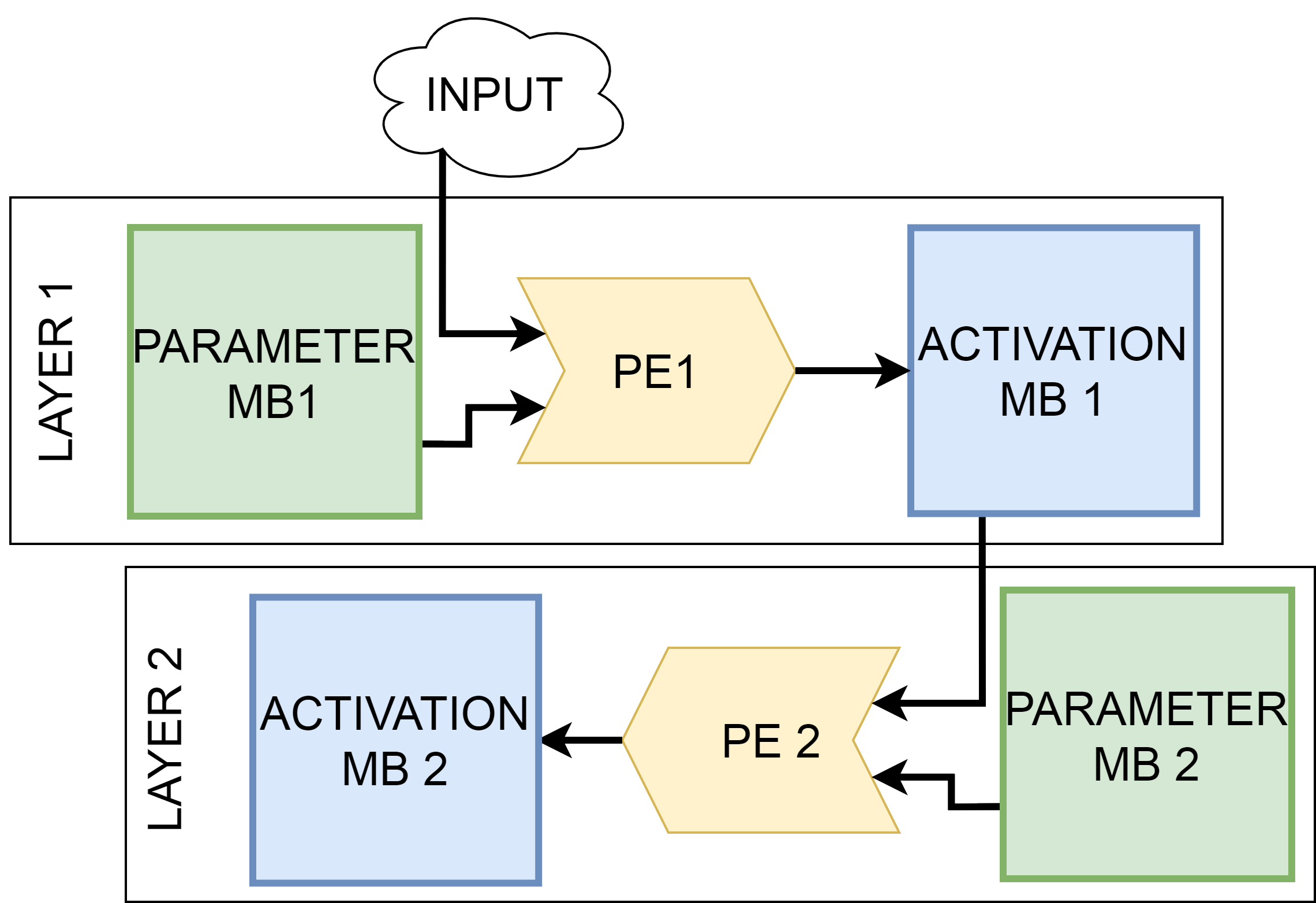}
\caption{DNN inference accelerator architecture with hybrid memory banks \textit{Parameter MB} and \textit{Activation MB} for storing DNN parameters and activation values respectively}
\label{design_arch}
\end{figure}
The architecture design of the Von-Neumann inference-only accelerator considered for this work is shown in Fig. \ref{design_arch} (only two layers shown). The accelerator uses hybrid memory banks (MB) to store both parameters (in parameters MB) and activations (in activation MB). Memory banks help improve the memory bandwidth by increasing the number of read channels. Also, the memory  to  compute-unit  distance  is  minimized which  improves  the  speed and  energy  efficiency  of  data-transfer \cite{ozturk2008ilp,koc2006minimizing}. The processing elements (PE) perform MAC operations between input and parameter data and store the results into activations MB which are used as inputs in the next layer. For each layer, the hybrid memory configurations may vary depending on the amount of surgical noise that needs to be introduced.
\section{Characterization of surgical noise}
\label{surgical_noise}
In hybrid memory architectures, at scaled voltages, bit-errors in the LSB 6T-SRAM cells introduce noise in the stored data values. To estimate the bit-error noise at various scaled voltages, we design a 6T-SRAM cell in 22nm technology using predictive models. The transistors are sized to have a nominal static read noise margin of 195 mV and write margin of 250 mV. Next we follow the procedure adopted by \cite{hybridmem} to calculate the bit-error rate values for different $V_{DD}$. Using the bit-error rates, we can model the bit-error noise in hybrid memories as a function of $V_{DD}$ and 8T-6T ratio $r$. 

Through our experiments we find that the mean bit-error noise exhibits some amount of deterministic behaviour depending upon the hybrid memory configurations: supply voltage $V_{DD}$ and 8T-6T ratio $r$. Note, $r$ implies $\frac{\#8T cells}{\#6T cells}$. Hence, we can deliberately introduce strategically crafted bit-error noise using appropriate hybrid memory configurations to improve the adversarial robustness of DNNs and adversarially attack hybrid memory architectures. We call this controlled bit-error noise as \textbf{surgical noise} and the process of deliberately introducing different amounts of surgical noise into specific sections of hybrid memories as \textbf{surgical noise injection (SNI)}.  

\begin{figure}[h!]
\centering
\includegraphics[width=0.35\textwidth]{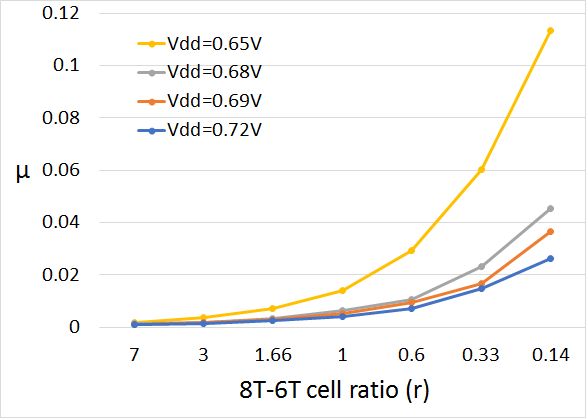}
\caption{Variation of average surgical noise perturbation $\mu$ with 8T-6T SRAM cell ratio $r$ for different supply voltages $V_{DD}$}
\label{sn_vs_r_vs_vdd}
\end{figure}

Mathematically, we can formulate the surgical noise by considering a vector $\nu_{l}$ being stored in layer $l$'s hybrid memory bank. Then, the vector after addition of surgical noise, $\nu_{sni, l}$, is given by the Equation \ref{N}.
\begin{equation}
    \label{N}
    \nu_{sni, l} = \nu_l + \mathcal{N}, ~where ~\mathcal{N} ~= ~f(V_{DD}, r)
\end{equation}
\begin{equation}
    \label{mue}
    \mu = mean(|\mathcal{N}|) = mean(|\nu_{sni, l}-\nu_l|) 
\end{equation}
Where $\mathcal{N}$ is the surgical noise added to the vector, $\mu$ is a scalar value shown in Equation \ref{mue} called the average surgical noise perturbation. In other words, $\mu$ denotes the net absolute perturbation introduced on vector $\nu_l$. The variation of $\mu$ for different values of $r$ and $V_{DD}$ is shown in Fig. \ref{sn_vs_r_vs_vdd}. It can be observed that with lower 8T-6T ratios, the value of $\mu$ increases because of the rise in the number of erroneous 6T-SRAM cells. Likewise, since the value of bit-error rates in 6T-SRAMs increase with higher voltage scaling, it can be seen that the overall $\mu$ value increases for lower $V_{DD}$ values.
\section{Improving Adversarial Robustness with Surgical Noise}
\label{adv_rob}
In this section, we investigate the idea of introducing surgical noise into hybrid memory architectures to improve the adversarial robustness of the model. 
\subsection{Methodology to determine the amount of surgical noise for each layer}
\begin{figure}[h!]
\centering
\includegraphics[width=0.4\textwidth]{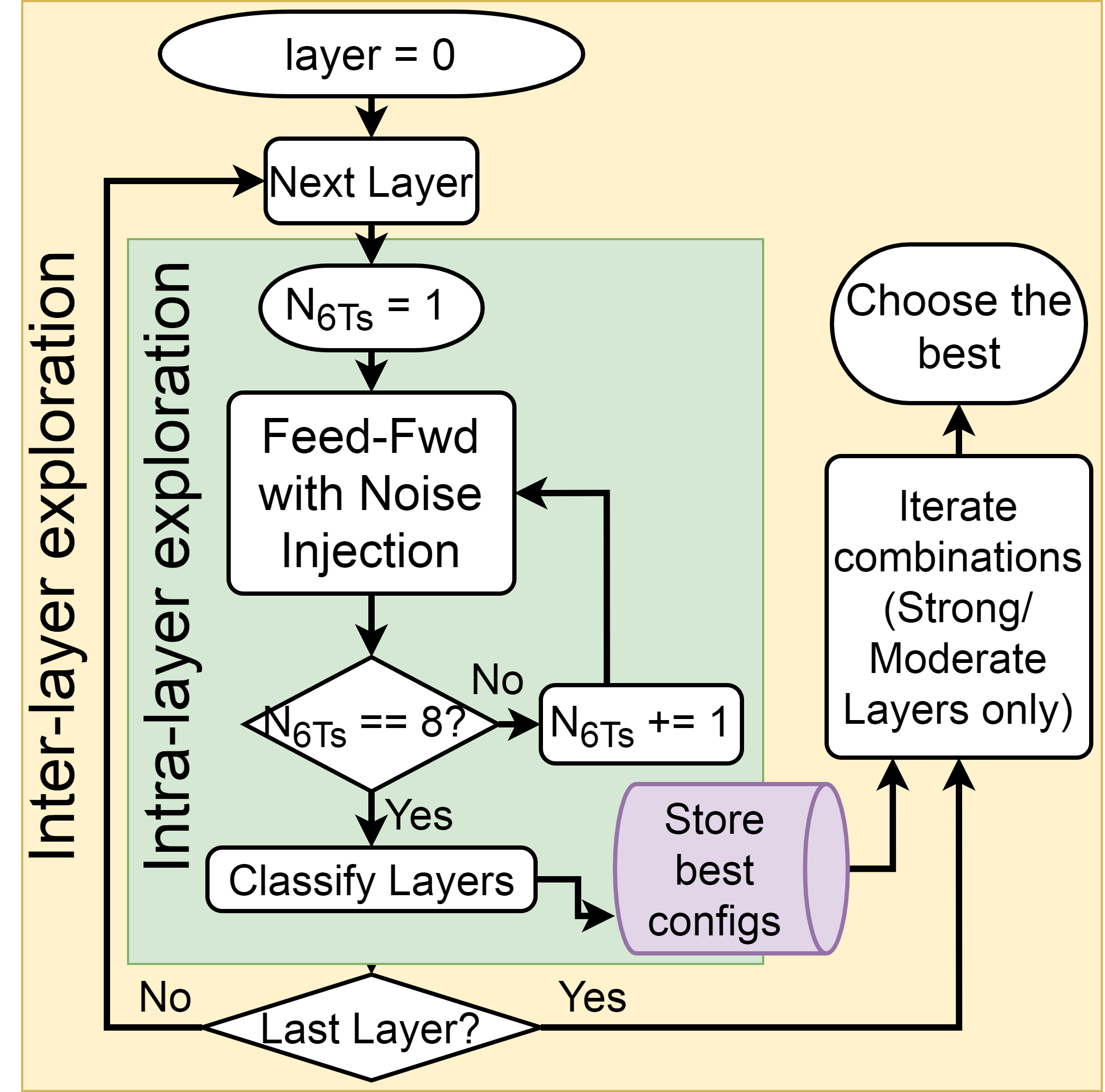}
\caption{Methodology for selecting the DNN layers suitable for SNI and determining required $\mu$ values for the suitable layers. The required $\mu$ is obtained via choosing the right hybrid memory configurations shown in Fig. \ref{sn_vs_r_vs_vdd}}
\label{algo_rob}
\end{figure}
To improve the adversarial robustness of the DNN model, we introduce surgical noise into the hybrid memory banks of specific layers of the DNN accelerator shown in Fig. \ref{design_arch}. We would like to point out that one can follow either or both the approaches- \textbf{1) Introducing surgical noise into the \textit{Parameter MB}, 2) Introducing surgical noise into the \textit{Activation MB}}. For both the approaches, the heuristics-based methodology proposed in Fig.\ref{algo_rob} can be applied. However, from our experiments we find that introducing surgical noise into the \textit{Acivation MB} yields much better result than introducing surgical noise into the \textit{Parameter MB}.

The heuristics-based methodology shown in Fig. \ref{algo_rob} determines the DNN layers suitable for SNI and the amount of surgical noise that needs to be introduced into the suitable layers to yield higher adversarial robustness. The amount of surgical noise is controlled by changing the hybrid memory configurations. To reduce the design space exploration however, we maintain the $V_{DD}$ at some constant value and vary the parameter $r$. The goal here is to check individual layers for their sensitivity towards SNI. 

To determine the suitable amount of $\mu$ required in each layer, the proposed methodology, in Fig. \ref{algo_rob} incrementally iterates over possible number of 8T-6T SRAM cell ratios $r$ (i,e \#6T-SRAM cells $\in [1,7]$) and introduces corresponding surgical noise into the layer. At each iteration, the DNN model is attacked using the FGSM method and the resulting adversarial accuracy is recorded. Note, while calculation of gradients for FGSM attack, surgical noise is not introduced into any of the DNN activations. Among all the iterations, the hybrid memory configuration that produced the best adversarial accuracy is stored for later evaluations. This process is repeated until all the layers are analysed. 

The DNN layers that produce more than 10\% higher adversarial accuracy upon SNI than the baseline models without introduction of surgical noise, are labeled as strong layers. Similarly, layers that produced higher than 5\% improvement in adversarial accuracy than baseline models are categorized as moderate layers. Weak response layers are the ones which produce same or lower adversarial accuracy upon SNI than the baseline models. At the end of this stage, the labels of individual DNN layers along with their favourable hybrid memory configurations are stored in the \textit{store best configs} module.

\begin{figure}[h!]
\includegraphics[width=0.5\textwidth]{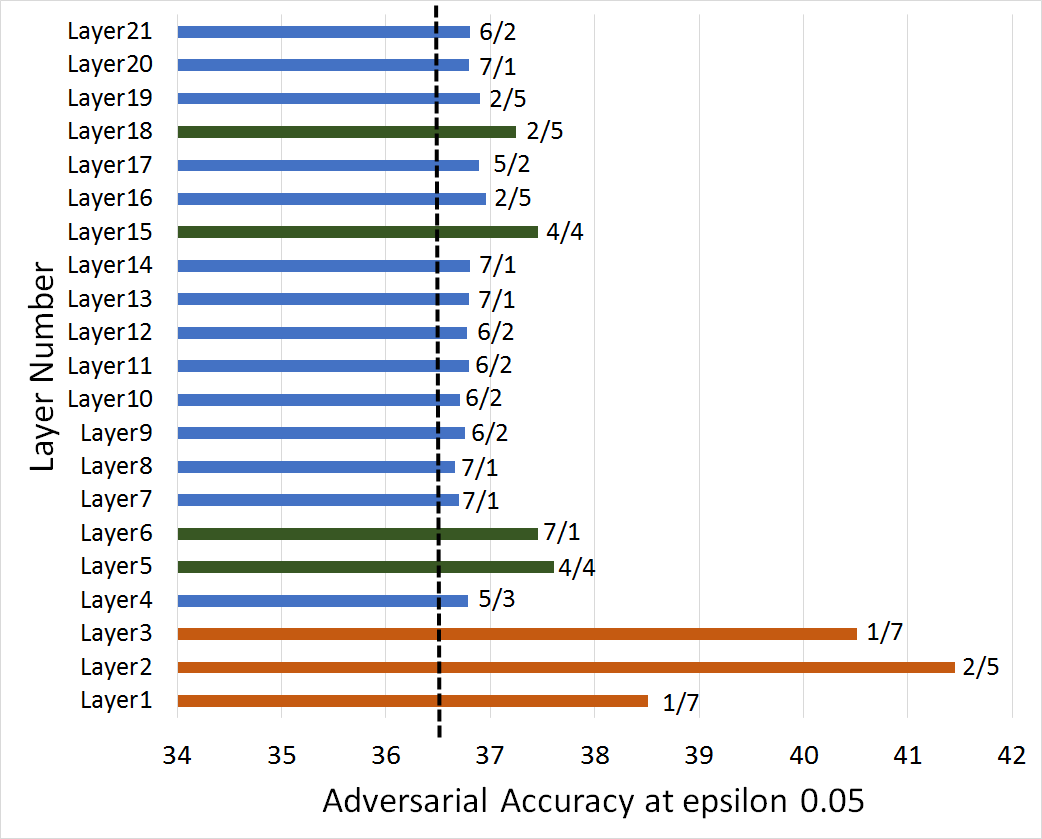}
\caption{Adversarial accuracy (with FGSM attack of strength $\epsilon = 0.05$) of the VGG19 network when surgical noise is introduced into each layer individually. For each layer, the value of $r$ corresponding to the best adversarial accuracy with SNI has been shown. The value of $V_{DD} = 0.68V$ }
\label{ind_layers}
\end{figure}

As an example, Fig. \ref{ind_layers} shows the \textit{store best configs} module data which comprises of the best adversarial accuracy obtained by introducing surgical noise into the corresponding layers individually, under FGSM ($\epsilon$= 0.05) attack. The 8T-6T ratios $r$ (mentioned alongside each bar) at $V_{DD}$= 0.68V are the corresponding favourable hybrid memory configurations for each layer. Strong layers shown in red produce high increment in adversarial accuracy while moderate (in green) and weak layers (in blue) show little or no increment respectively. For reference, the black dotted line shows the adversarial accuracy of the baseline model- without any SNI.

After categorizing the layers, the proposed methodology tries to find the best possible combination of layers that yield the best adversarial robustness. For this the strong and moderate layers with their favourable configurations (i,e, with configuration that generated good accuracy when tested individually) are combined and the adversarial accuracy of the DNN model is recorded under FGSM attack. For example, from the Fig.\ref{ind_layers}, we see that layers [1, 2, 3] are strong while [5, 6, 15, 18] are moderate layers with favourable 8T-6T ratios as (3-5), (2-6), (5-3) and so on. The proposed methodology iterates over various combinations of the strong and moderate layers, ie \{[1,2], [1,5], [2,3], [1,2,3,6] and so on\}, with their respective favourable 8T-6T ratios and chooses the combination that yields the best adversarial accuracy. This gives us the final set of layers along with their hybrid memory configurations that yield the best adversarial robustness.
\subsection{Possible hybrid memory architectures}

Table \ref{layer_config vgg} and Table \ref{layer_config resnet} show the 8T-6T ratios (at $V_{DD} = 0.68$V) for individual layers of VGG19 and ResNet18 DNN architectures trained to perform classifications on benchmark datasets- CIFAR10 and CIFAR100. The last column of each table shows the DNN's clean accuracy with SNI into the respective layers and the deviation with respect to clean accuracy of the baseline model (with no SNI). In the Table \ref{layer_config vgg}, layers with P denote the pooling layers while in Table \ref{layer_config resnet}, S denotes the residual layers in the ResNet18 network. The layers chosen using our proposed methodology above are shown with the respective $r$ values of their hybrid activation memories. Note that we do not perform SNI on the DNN parameter memories. The rest of the layers in which introducing surgical noise into the hybrid activation memories will have a degrading effect on the DNN's performance are shown using H. Meaning that they are composed of either homogeneous 8T-SRAM or homogeneous 6T-SRAM memory. 

The choice of memory to design the non-noisy layers gives rise to two design paradigms: \textbf{1) Design for Power efficiency 2) Design for Area efficiency.}  
\begin{table*}[]
\centering
\caption{Layer wise activation memory configurations with the corresponding 8T-6T values for the VGG19 network. P denotes the pooling layers. H denotes homogeneous activation memory hence no 8T-6T ratios have been mentioned}
\label{layer_config vgg}
\resizebox{\textwidth}{!}{%
\begin{tabular}{|c|c|c|c|c|c|c|c|c|c|c|c|c|c|c|c|c|c|c|c|c|c|c|c|}
\hline
Layer                         & 0 & 1   & \begin{tabular}[c]{@{}c@{}}2 \\ (P)\end{tabular} & 3 & 4   & \begin{tabular}[c]{@{}c@{}}5\\ (P)\end{tabular} & 6 & 7 & 8 & 9 & \begin{tabular}[c]{@{}c@{}}10\\ (P)\end{tabular} & 11 & 12 & 13 & 14 & \begin{tabular}[c]{@{}c@{}}15\\ (P)\end{tabular} & 16 & 17 & 18 & 19 & \begin{tabular}[c]{@{}c@{}}20\\ (P)\end{tabular} & $V_{DD}$ & \begin{tabular}[c]{@{}c@{}}Clean Accuracy/Deviation\end{tabular} \\ \hline
\multicolumn{1}{|l|}{CIFAR10} & H & 3/5 & 2/6                                              & H & 5/3 & H                                               & H & H & H & H & H                                               & H & H & H & H & H                                                & H & H & H & H & H                                                & 0.68V    & 88.78 / 2.61                                                       \\ \hline
CIFAR100                      & H & 3/5 & 2/6                                              & H & H   & 5/3                                             & H & H & H & H & H                                                & H & H & H & H & H                                                & H & H & H & H & H                                                & 0.68V    & 67.3 / 2.9                                                        \\ \hline
\end{tabular}%
}
\end{table*}
\begin{table*}[]
\centering
\caption{Layer wise activation memory configurations with the corresponding 8T-6T values for the ResNet18 network. S denotes the shortcut layers. H denotes homogeneous activation memory hence no 8T-6T ratios have been mentioned}
\label{layer_config resnet}
\resizebox{\textwidth}{!}{%
\begin{tabular}{|c|c|c|c|c|c|c|c|c|c|c|c|c|c|c|c|c|c|c|c|c|c|c|c|c|c|c|}
\hline
Layer                         & 0 & 1   & \begin{tabular}[c]{@{}c@{}}2 \\ (S)\end{tabular} & 3 & 4   & \begin{tabular}[c]{@{}c@{}}5\\ (S)\end{tabular} & 6 & 7 & \begin{tabular}[c]{@{}c@{}}8\\ (S)\end{tabular} & 9 & 10 & \begin{tabular}[c]{@{}c@{}}11\\ (S)\end{tabular} & 12 & 13 & \begin{tabular}[c]{@{}c@{}}14\\ (S)\end{tabular} & 15 & 16 & \begin{tabular}[c]{@{}c@{}}17\\ (S)\end{tabular} & 18 & 19 & \begin{tabular}[c]{@{}c@{}}20\\ (S)\end{tabular} & 21 & 22 & \begin{tabular}[c]{@{}c@{}}23\\ (S)\end{tabular} & $V_{DD}$ & \begin{tabular}[c]{@{}c@{}}Clean Accuracy/Deviation\end{tabular} \\ \hline
\multicolumn{1}{|l|}{CIFAR10} & H & 4/4 & 5/3                                              & H & 6/2 & H                                               & H & H & H & H & H                                                & H & H & H & H & H                                                 & H & H & H & H & H  & H & H & H                                               & 0.68V    & 89.2 / 6.14                                                       \\ \hline
CIFAR100                      & 5/3 & H & 6/2                                              & 6/2 & H & H                                             & H & H & H & H & H                                                & H  & H  & H  & H  & H                                                & H  & H  & H  & H  & H & H & H & H                                               & 0.68V    & 69.4 / 7.1                                                        \\ \hline
\end{tabular}%
}
\end{table*}
\subsubsection{Design for Energy Efficiency}
\begin{figure}[h!]
\centering
\includegraphics[width=0.4\textwidth]{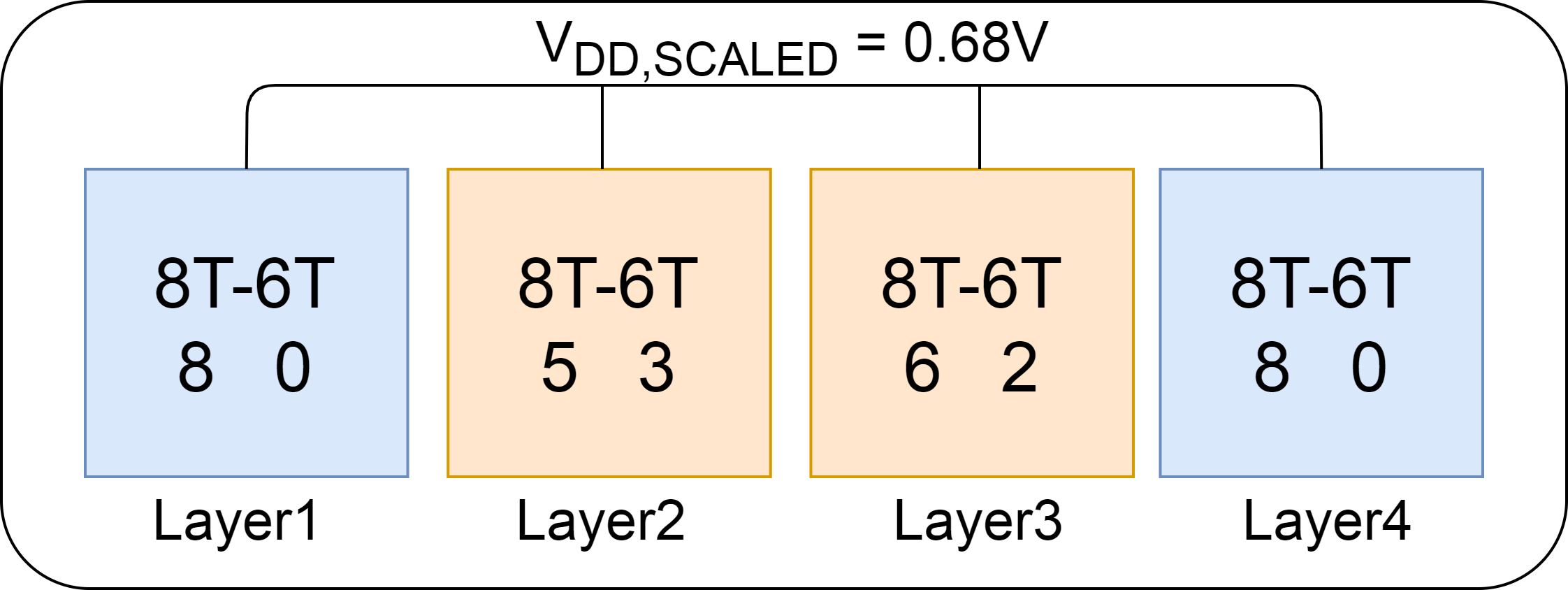}
\caption{Activation memory architecture design for energy efficient implementation. Both hybrid memory banks (red) and homogeneous memory banks (blue) are connected to the same scaled voltage line.}
\label{pe_hm}
\end{figure}
In this design, the activation memories of the non-noisy DNN layers are stored in homogeneous 8T-SRAM memory. Fig.\ref{pe_hm} shows a block design for this approach where the red blocks represent the noisy activation memories and the blue blocks are the non-noisy activation memories. Here, a common supply voltage $V_{DD, SCALED}$ which is a scaled voltage, is fed to all the DNN layers. The result is that the overall energy is lowered because of voltage scaling. Since 8T-SRAM cells are error resilient at low voltages, scaling the supply voltage will have no effect on the DNN model accuracy. 
\subsubsection{Design for Area Efficiency}
\begin{figure}[h!]
\centering
\includegraphics[width=0.4\textwidth]{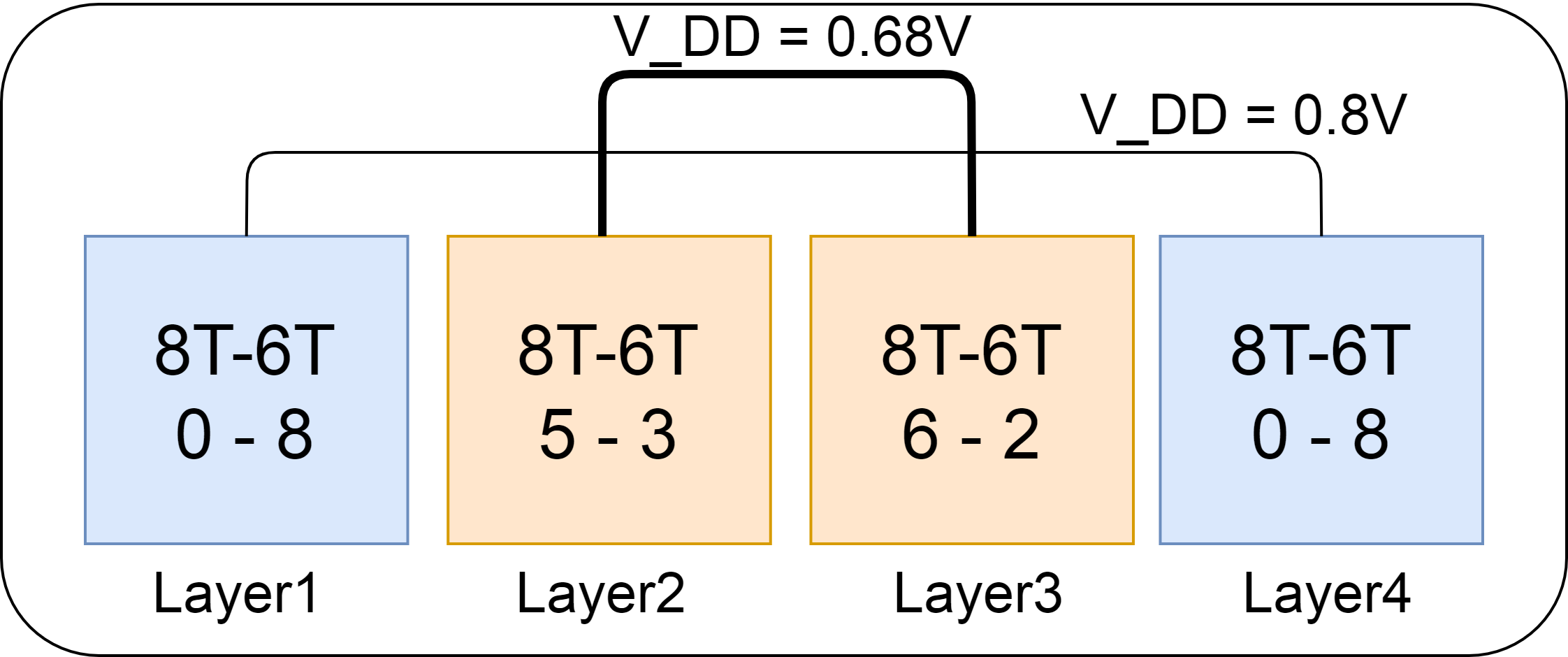}
\caption{Activation memory architecture design for area efficient implementation. Here hybrid memory banks (red) are connected to the scaled voltage line and homogeneous memory banks (blue) are connected to the nominal voltage line.}
\label{ae_hm}
\end{figure}
\begin{figure}[h!]
\centering
\includegraphics[width=0.4\textwidth]{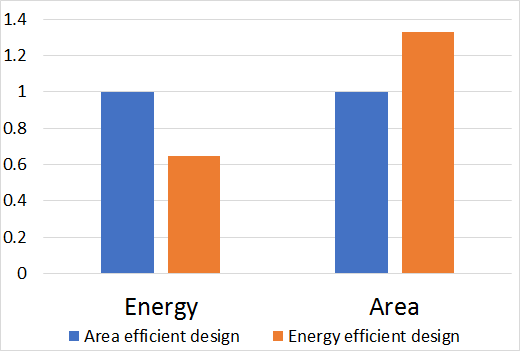}
\caption{Comparative analysis of the energy and area efficient design paradigms}
\label{pe_ae}
\end{figure}
In contrast to the previous approach, if the non-noisy layers are designed using homogeneous 6T-SRAM cells, the silicon footprint will be greatly reduced. However, if we supply the same $V_{DD,SCALED}$ value to all the layers, like in the previous case, bit-error noise in the homogeneous 6T-SRAM memories will degrade the DNN's performance significantly. Hence, in order to maintain the DNN's performance, two supply voltage lines, one with $V_{DD}$, the nominal voltage value is fed to the homogeneous 6T-SRAM memories while the other with $V_{DD,SCALED}$ value is fed to the chosen noisy layers. This has been shown in Fig.\ref{ae_hm}. 

A comparative analysis shown in Fig.\ref{pe_ae}, explains the energy and area advantages of individual design paradigms. The homogeneous 8T-SRAM memory design is energy efficient as it consumes around 35.45\% less energy than when the activations are stored in 6T-SRAM memory. However, this design paradigm incurs a 30\% increase in silicon area than homogeneous 6T-SRAM memory architecture.
\subsection{Results and Discussions}
\subsubsection{Experimental Setup}
The surgical noise values are estimated using SPICE simulations with 22nm predictive models. For all our experiments, to estimate the adversarial accuracy values with and without SNI, the DNN models are implemented on a DNN functional simulator designed using Pytorch. The functional simulator provides support for introducing specific amounts of surgical noise as required into specific layers of the DNN model. For faster simulations, we implement the functional simulator on a 2 Nvidia RTX2080ti GPU back-end platform. Additionally, in all our experiments, the DNNs are trained with 8-bit homogeneous bit-precision.
\subsubsection{Datasets and Network Architectures}
\label{section_data_network}
We use two visual datasets namely CIFAR10 and CIFAR100. The CIFAR10 dataset has 60K examples distributed among 10 classes. The CIFAR100 dataset is similar to CIFAR10 with the difference being that there are 100 classes. Both have 50K training and 10K test examples. The reduction in number of examples per class in CIFAR100 leads to lower clean and adversarial accuracies.

For both the datasets, we evaluate our results using two different network architectures- 1) VGG19 network considered in this work has 16 convolutional layers with 5 pooling layers in between and 1 fully connected layer at the end of the network. 2) The ResNet18 network which is a standard residual network consists of 4 blocks with each block containing 4 convolutional layers and 2 residual layer. The network has an average pooling layer and a fully connected layer at the end. We perform inferences with surgical noise on both datasets on two different types of network architectures to get an understanding of how different networks respond to noise injection and their adversarial performance.
\subsubsection{Results for CIFAR10 and CIFAR100 datasets}
To validate the performance of our proposed methodology for improving the adversarial robustness of DNNs using surgical noise, we employ white-box attacks, specifically FGSM attacks explained in the section \ref{adversarial_attacks}. To generate adversarial examples using the FGSM attacks, we perturb the input examples with varying strengths, $\epsilon$ between 0.05 to 0.3. Increasing the perturbation beyond 0.3 does not ensure "visibly minor changes" in the input images and hence makes no sense. Also we ensure that surgical noise is not introduced into the activations during the gradient calculation for the FGSM attack. However, during inference, we introduce surgical noise into the specific layers of the DNN according to Table \ref{layer_config vgg} and table \ref{layer_config resnet} chosen using our proposed methodology.
\begin{figure}[h!]
\centering
\begin{subfigure} {0.4\textwidth}
\includegraphics[width=\textwidth]{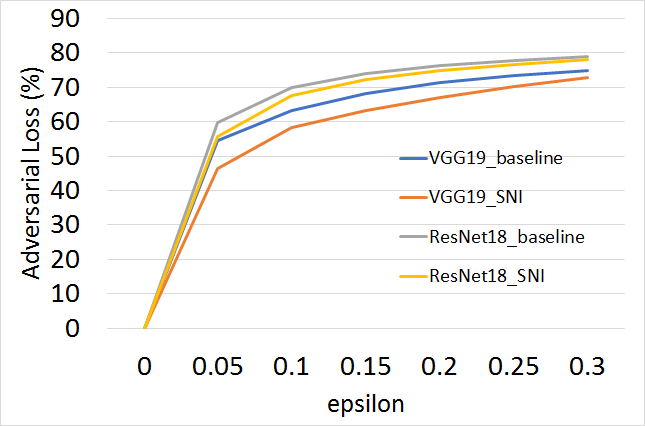}
\caption{}
\label{fig cifar10 r_v}
\end{subfigure}
\begin{subfigure} {0.4\textwidth}
\includegraphics[width=\textwidth]{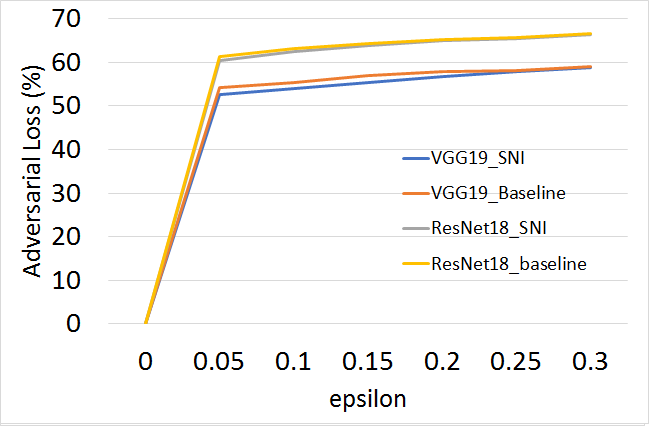}
\caption{}
\label{fig cifar100}
\end{subfigure}
\caption{Variation of adversarial loss with FGSM perturbation strength $\epsilon$ for a) CIFAR10 and b) CIFAR100 with SNI on both VGG19 and ResNet18 network architectures. Adversarial loss for \textit{Baseline} models is calculated without SNI into the DNN network.}
\end{figure}

Fig.\ref{fig cifar10 r_v} and Fig.\ref{fig cifar100} show the improvements in adversarial loss for CIFAR10 and CIFAR100 datasets respectively. For each dataset, we perform our experiments using both the VGG19 and ResNet18 network architectures. Adversarial loss is calculated using the Equation \ref{adv_loss}:
\begin{equation}
    adv~loss = clean~accuracy-adv~accuracy
    \label{adv_loss}
\end{equation}
Both clean and adversarial accuracy have been defined in section \ref{backgroud}. To compute the baseline adversarial loss, the clean accuracy of the DNN model without any SNI and the adversarial accuracy of the DNN model without SNI are computed corresponding to each $\epsilon$. Similarly, the adversarial loss of the DNN with surgical noise is computed using clean accuracy and adversarial accuracy of the DNN with SNI.

An important thing to note here is the clean accuracy of the DNN with SNI when compared to the clean accuracy of the baseline model. Table \ref{layer_config vgg} and Table \ref{layer_config resnet} show the clean accuracy of the DNN models after SNI along with the deviation from the clean accuracy of the baseline model. It can be observed that due to SNI, the clean accuracy value decreases slightly with respect to the baseline model. This accuracy degradation can be improved by re-training the surgical noise injected DNN model on clean examples. 
\subsubsection{Discussions}
It is evident from Fig.\ref{fig cifar10 r_v} and Fig.\ref{fig cifar100} that SNI into hybrid memories storing the activations of the DNN leads to a decrease in adversarial loss. Adversarial loss is the difference between the \textit{clean accuracy} and the \textit{adversarial accuracy} of the DNN model. It should be noted that lower value of adversarial loss means higher robustness. Interestingly, we find that initial layers in both the network architectures offer higher adversarial robustness with SNI. This is expected since, in the later layers, features become more and more specific and can be easily corrupted by SNI. Additionally, it must be noted that the adversarial loss for the ResNet18 network is higher than the VGG19 network for both CIFAR10 and CIFAR100 datasets. Also, the improvements in adversarial loss is less for residual networks than VGG networks.

At this point, we also want to highlight the challenge of effective design space exploration in our proposed methodology. The multitude of layers and hybrid memory configurations lead to a very high design space for optimization. Although, in our heuristics algorithm, we explored a small design space and showed that higher adversarial accuracy can be achieved by introducing surgical noise into hybrid memories, these results might certainly not be the most optimum. Employing more structured optimization algorithms like Genetic Algorithm \cite{whitley1994genetic} and Particle Swarm Optimization \cite{kennedy1995particle} might lead to much better results.

\section{Using surgical noise to attack hybrid memories}
\label{attack}
In this section, we bring out a potential vulnerability of hybrid memory based DNN accelerators by proposing a novel, white-box DNN parameter attack methodology. Unlike the previous section, where we used surgical noise in hybrid memories to improve the adversarial robustness of the DNN model, here we show that surgical noise can be strategically injected into a small, specific section of the hybrid memory storing the DNN parameters to cause high confidence misclassifications. We further show that this attack is adversarial in nature.
\subsection{Surgical noise based white-box attack on DNN parameters}
To design the white-box DNN parameter attack, we perform a gradient-based perturbation like the one used in FGSM \cite{goodfellow2014explaining}. In this work however, unlike FGSM, that perturbs the input image, we use the gradient-based method to perturb the DNN parameters for performing adversarial attacks. This has been shown in Equation \ref{adv_attack}. First, the gradients of the parameters with respect to the DNN model loss is calculated. Note, that during the gradient calculations with respect to the loss, no surgical noise is introduced into the layers of the DNN model. We define a variable $\epsilon$, that accounts for the magnitude of perturbation introduced into the DNN parameters. To introduce adversarial perturbations however, we need to ensure that the perturbation is applied along the direction of the gradients.
\begin{equation}
    \label{adv_attack}
    W_{adv} = W_{orig} + \epsilon ~ sign(\frac{\partial \mathcal{L}}{\partial W_{orig}})
\end{equation}where, $\mathcal{L}$ is the loss of the DNN model without adversarial attack and $W_{orig}$ is the original DNN parameter vector that is adversarially perturbed to create $W_{adv}$.

This method is fairly straightforward to execute in software. However, in hardware since surgical noise is crafted using bit-error noise due to the erroneous 6T-SRAM cells, we have to account for the probabilistic nature of perturbations introduced into the DNN parameters by surgical noise.

Hence, we propose a probabilistic approach for introducing adversarial perturbations into the DNN parameters stored in the hybrid memory. For this, we redefine the terms $\epsilon$ and $sign(\frac{\partial \mathcal{L}}{\partial W_{orig}})$ as $\mu$ and $\mathcal{D}$ respectively. The term $\mu$ shown in Equation \ref{mu} is same as the one shown in Equation \ref{mue}. $\mu$ is the average magnitude of the surgical noise $\mathcal{N}_{W_{orig}}$ for all the elements in the vector $W_{orig}$. Similarly, $\mathcal{D}$ shown in Equation \ref{D} denotes the direction of perturbation due to surgical noise for each element in $W_{orig}$. It must be noted that the terms $\mu$ and $\mathcal{D}$ are equivalent to the terms $\epsilon$ and $sign(\frac{\partial \mathcal{L}}{\partial W_{orig}})$ in Equation \ref{adv_attack} respectively.
\begin{equation}
    \label{mu}
     \mu = mean((|\mathcal{N}_{W_{orig}}|)
\end{equation}
\begin{equation}
    \label{D}
     \mathcal{D} = sign(\mathcal{N}_{W_{orig}})
\end{equation}
\begin{equation}
    \label{sni_adv}
    W_{adv,sni} = W_{orig} + \mu \times \mathcal{D}
\end{equation}
Hence, Equation \ref{adv_attack} can be reformulated to realise surgical noise-based adversarial attack on DNN parameters $W_{orig}$ as shown in Equation \ref{sni_adv}. Note that $\mu$ is a scalar and $\mathcal{D}$ is a vector having the same dimensions as $W_{orig}$. Since, the terms $\mathcal{D}$ and $sign(\frac{\partial Loss}{\partial W_{orig}})$ are equivalent, we want their values to closely resemble each other for introducing an adversarial attack on DNN parameters stored in the hybrid memory. 
\subsection{Proposed methodology for the surgical noise based adversarial attack}
Due to the probabilistic nature of surgical noise in hybrid memories, it is not always guaranteed that the perturbations in each and every DNN parameter $W$ ($W$ denotes the vector containing the DNN parameters) due to surgical noise follow the same direction as $\frac{\partial \mathcal{L}}{\partial W}$. To overcome this issue, we propose a methodology that perturbs a smaller section of the hybrid memory $W_{i} \subset W$ for which the surgical noise induced perturbations follow the same direction as the gradient of loss with respect to the DNN parameters stored in that section. In other words, we choose a section $W_i \subset W$ such that $\mathcal{D}_i$ is very close to $sign(\frac{\partial \mathcal{L}}{\partial W_{i}})$. 

\begin{figure}[h!]
\centering
\includegraphics[width=0.4\textwidth]{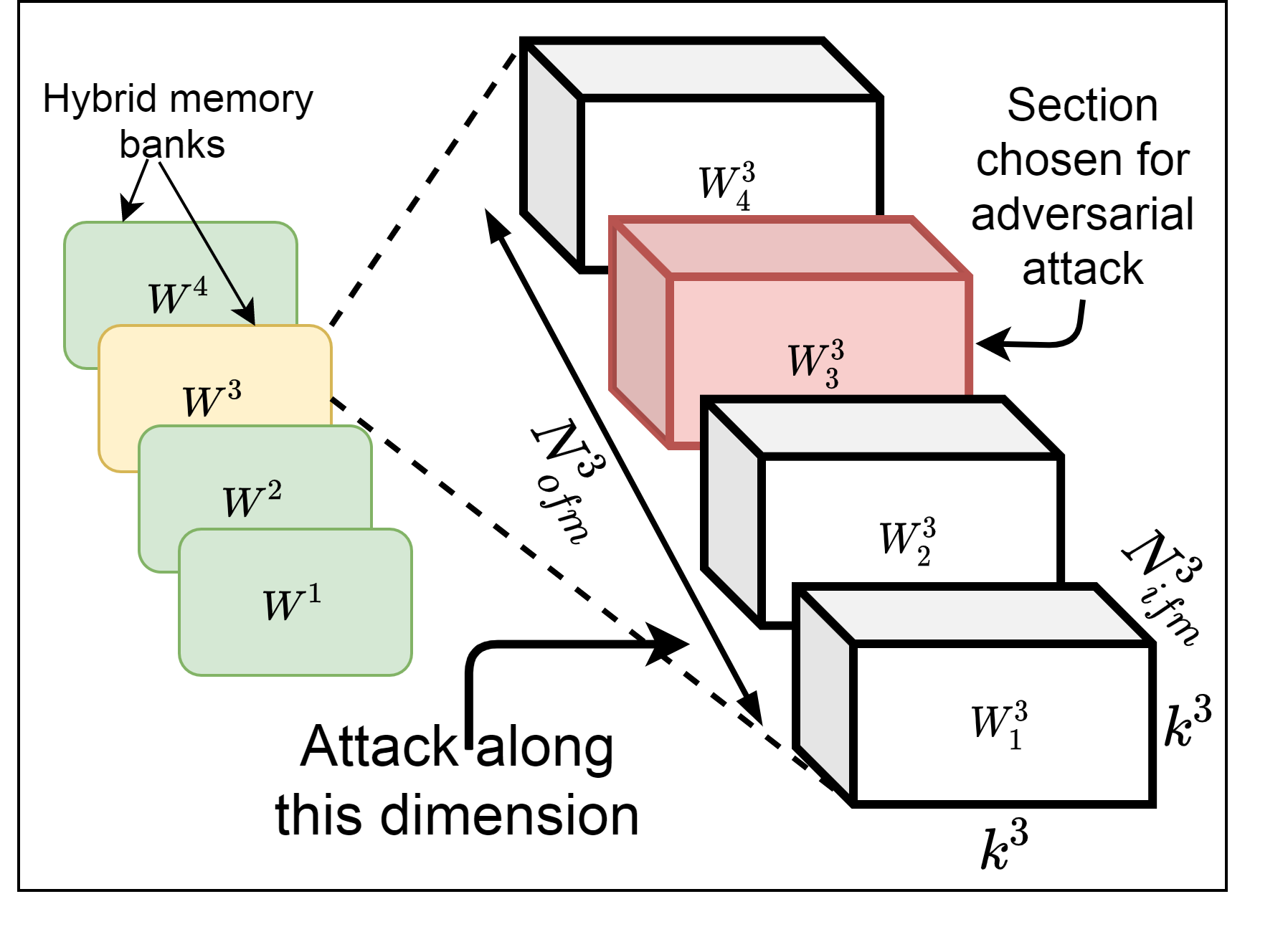}
\caption{Methodology to select a section from a hybrid memory bank to introduce adversarial perturbations using surgical noise}
\label{mem_bank_Arch}
\end{figure}

Fig. \ref{mem_bank_Arch} shows how the surgical noise-based attack is performed on the DNN parameters. We assume that the parameters of each layer in the model are stored in separate hybrid memory banks. For each layer $l$, the corresponding hybrid memory bank stores its parameters in the form of a vector $W^l$. The dimensions of this vector is $[N^l_{ofm},N^l_{ifm},k^l,k^l]$, where $N^l_{ofm}$ is the number of output feature maps of the layer $l$, $N^l_{ifm}$, the number of input feature maps and $k^l$ is the kernel dimension. For this work, $W^l_{i}s$ are defined as vectors along the first dimension ($N^l_{ofm}$) of $W^l$. In other words, we perform DNN parameter attack on a smaller section $W^l_i$ of dimensions $[N^l_{ifm},k^l,k^l]$.

\begin{figure}[h!]
\centering
\includegraphics[width=0.3\textwidth]{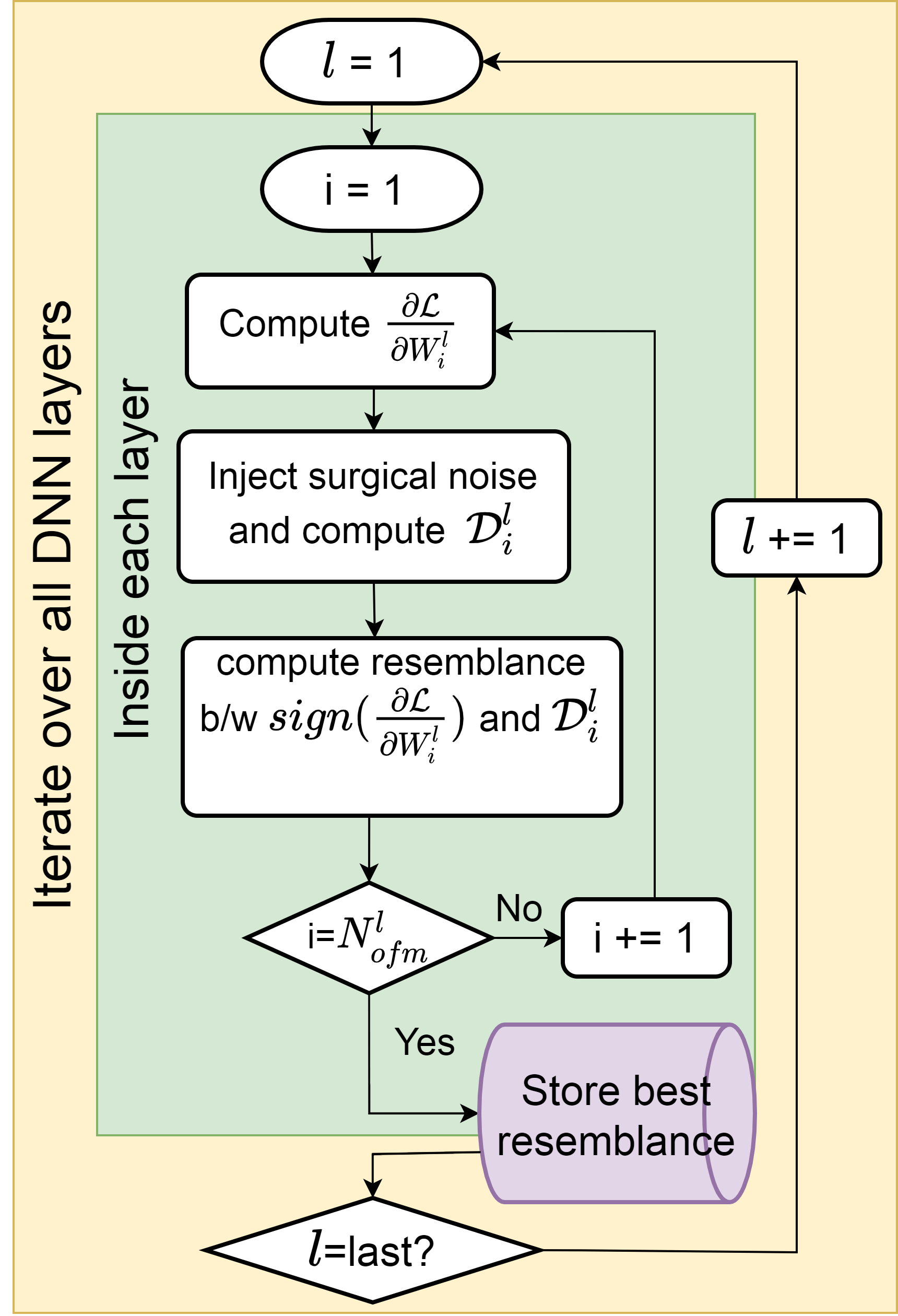}
\caption{Proposed methodology for selecting the appropriate section $W^l_i$ from $W$ to ensure adversarial DNN parameter attack.}
\label{wa_algo}
\end{figure}

We propose a methodology shown in Fig. \ref{wa_algo} to find the best section $W^l_i$ in each layer $l$ for introducing surgical noise-based adversarial attack. First, based on the average magnitude of surgical noise perturbation $\mu$ desired, we choose the hybrid memory configurations from the data shown in Fig. \ref{sn_vs_r_vs_vdd}. Then, for each layer, we iterate over all the $W^l_i$s in the vector $W^l$ along the $N_{ofm}$ dimension. For each $W_i^l$, we calculate the value of $sign(\frac{\partial \mathcal{L}}{\partial W^l_{i}})$. Next, for the section $W^l_i$, we introduce surgical noise into the DNN parameters stored in the section and calculate the value of $\mathcal{D}^l_i$. After this, a comparison is made between $\mathcal{D}^l_i$ and $sign(\frac{\partial \mathcal{L}}{\partial W^l_{i}})$ to check the percentage match between the two vectors. This is repeated for each and every section $W^l_i$ and the section with the highest match value is chosen for the attack.

\begin{figure*}[t!]
\centering
\begin{subfigure} {\textwidth}
\includegraphics[width=0.95\textwidth]{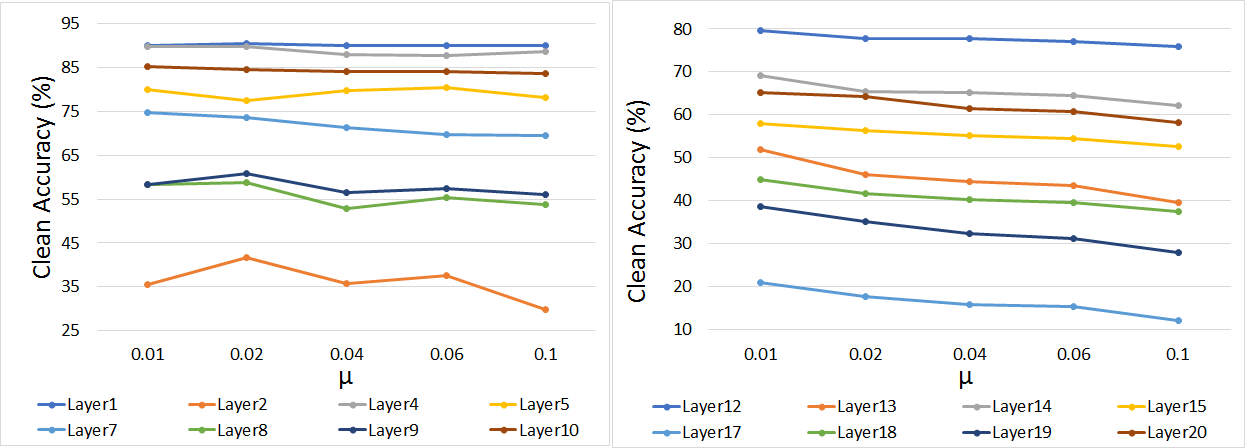}
\caption{}
\label{vgg19_wa}
\end{subfigure}

\begin{subfigure} {\textwidth}
\includegraphics[width=0.95\textwidth]{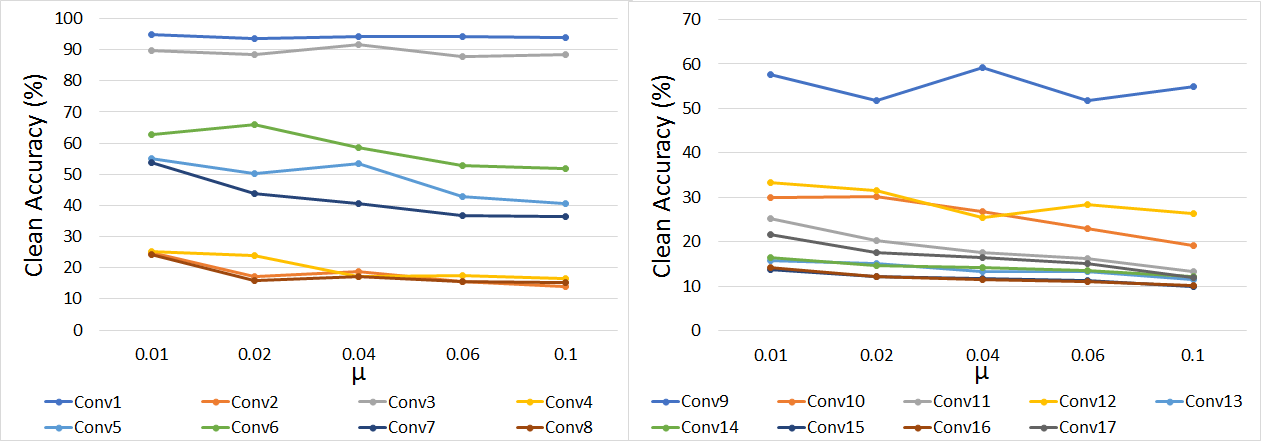}
\caption{}
\label{resnet_wa}
\end{subfigure}
\caption{Plots showing variation of clean accuracy of a)VGG19 b)RESNET18 networks, on CIFAR10 dataset for different values of surgical noise induced perturbations $\mu$}
\end{figure*}
\begin{table}[h!]
\centering
\caption{Table showing the values of $\mu$ corresponding to various hybrid memory configurations ($V_{DD}$, $r$) for performing surgical noise-based attacks}
\label{perturbation_table}
\resizebox{0.3\textwidth}{!}{%
\begin{tabular}{|c|c|c|}
\hline
$\mu$ & $8T-6T$ & $V_{DD}$ \\ \hline
0.01           & 3-5 & 0.68V                         \\ \hline
0.02           & 1-7 & 0.72V                         \\ \hline
0.04           & 1-7 & 0.69V                         \\ \hline
0.06           & 2-6 & 0.65V                         \\ \hline
0.1           & 2-6 & 0.65V                         \\ \hline
\end{tabular}%
}
\end{table}
\section{Results}
\subsection{Experimental Setup}
To demonstrate the surgical noise-based adversarial attack on the DNN parameters, we perform experiments on two DNN architectures VGG19 and RESNET18. To compare their sensitivity towards the attack, both have been trained on a single visual dataset- CIFAR10. A detailed account of the dataset and network architectures have been presented in section \ref{section_data_network}. The baseline VGG19 and ResNet18 networks without surgical noise-based attack, yield a clean accuracy of 91.39\% and 95.34\%, respectively, for the CIFAR10 dataset. Also, note that we do not attack the residual layers in the ResNet18 network to maintain simplicity. For all our experiments, we implement our proposed method using our DNN Functional simulator.
\subsection{Results of surgical noise-based DNN parameter attack}
Based on our proposed methodology of attacking the DNN parameters stored in the hybrid memory, we report the accuracy of the DNN model under various surgical noise perturbation strengths $\mu$. The required value of $\mu$ is obtained from Fig. \ref{sn_vs_r_vs_vdd} and have been shown in Table \ref{perturbation_table}. For each perturbation strength $\mu$, we introduce the respective surgical noise into the section of hybrid memory chosen by our proposed methodology. 

The demonstration of our analysis has been shown in Fig. \ref{vgg19_wa} and Fig. \ref{resnet_wa}. The analysis clearly shows that different layers have different levels of sensitivity towards the surgical noise-based adversarial attack. For all the classification experiments shown, the network classifies the images with a confidence score of more than 90\%. The average percentage match between $\mathcal{D}^l_i$ and $sign(\frac{\partial \mathcal{L}}{\partial W^l_{i}})$ for the chosen section $i$ of the DNN layer $l$ is above 99\%. Thus, we ensure the surgical noise-based DNN parameter attack to be adversarial in nature.

\subsection{Discussions}
We see that the input layer in both the networks are considerably robust even for high $\mu$ values. Because the input layers are tasked with coarse feature extraction, introducing any perturbations into their synaptic weights will have little or no effects on the output. However, we see that the very next layer (layer 2) for both VGG19 and ResNet18 networks pose a strong vulnerability. Here, even for low $\mu$ values, the DNN clean accuracy starts degrading by over 60\%. Additionally, a contrasting difference in the susceptibility towards surgical noise-based attack that can be noticed between the two DNN architectures is that VGG19 has more number of robust layers than ResNet18. 

\begin{figure}[h!]
\includegraphics[width=0.5\textwidth]{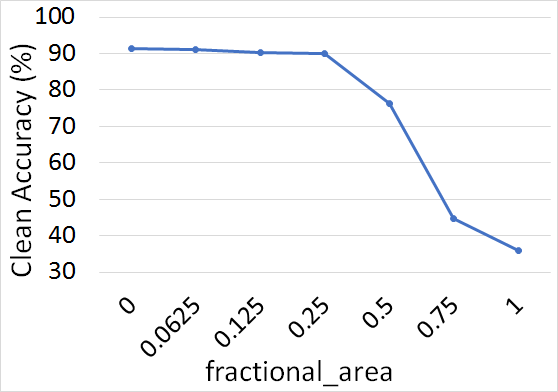}
\caption{Variation of clean accuracy of the VGG19 network when smaller subsections of the selected section $W^l_i$ from layer 2 is attacked}
\label{finer}
\end{figure}

We perform another analysis to check if the DNN's clean accuracy is affected by introducing surgical noise into sub-sections of size smaller than $W^l_i$ of the hybrid memory. For this, the size of the sub-section in the given section $W^l_i$ is varied from a quarter to three-fourths of the size of the section. This has been shown in Fig. \ref{finer}. In this example, we choose the selected section of the second layer in the VGG19 network and gradually increment the size of the sub-section while introducing surgical noise perturbation of magnitude $\mu$= 0.01 in order to check the accuracy and confidence value of the DNN. We observe that even for a smaller attack region, our surgical noise-based adversarial attack can successfully degrade the DNN's performance.

\section{Conclusion}
In this work we show that bit-error noise in hybrid memories due to erroneus 6T-SRAM cells can be controlled using the hybrid memory configurations. The surgical noise can be strategically injected into hybrid memories storing the activations and DNN parameters to improve the adversarial robustness of the DNN model. Additionally, the DNN parameters stored in the hybrid memory banks can be adversarially attacked using surgical noise injected in a strategic manner causing a huge degradation in the DNN inference accuracy. Hence, implementing 6T-8T hybrid memories in DNN hardware accelerators can not only make the design more energy efficient but can also impact the adversarial robustness of the DNN model.

\section{Acknowledgement}
This work was supported in part by National Science Foundation (Grant\#1947826), the Technological Innovation Institute, and the Amazon Research Award.
\ifCLASSOPTIONcaptionsoff
  \newpage
\fi



%
\bibliographystyle{IEEEtran}
\bibliography{bibli.bib}

\begin{thebibliography}{10}
\providecommand{\url}[1]{#1}
\csname url@samestyle\endcsname
\providecommand{\newblock}{\relax}
\providecommand{\bibinfo}[2]{#2}
\providecommand{\BIBentrySTDinterwordspacing}{\spaceskip=0pt\relax}
\providecommand{\BIBentryALTinterwordstretchfactor}{4}
\providecommand{\BIBentryALTinterwordspacing}{\spaceskip=\fontdimen2\font plus
\BIBentryALTinterwordstretchfactor\fontdimen3\font minus
  \fontdimen4\font\relax}
\providecommand{\BIBforeignlanguage}[2]{{%
\expandafter\ifx\csname l@#1\endcsname\relax
\typeout{** WARNING: IEEEtran.bst: No hyphenation pattern has been}%
\typeout{** loaded for the language `#1'. Using the pattern for}%
\typeout{** the default language instead.}%
\else
\language=\csname l@#1\endcsname
\fi
#2}}
\providecommand{\BIBdecl}{\relax}
\BIBdecl

\bibitem{szegedy2013intriguing}
C.~Szegedy, W.~Zaremba, I.~Sutskever, J.~Bruna, D.~Erhan, I.~Goodfellow, and
  R.~Fergus, ``Intriguing properties of neural networks,'' \emph{arXiv preprint
  arXiv:1312.6199}, 2013.

\bibitem{goodfellow2014explaining}
I.~J. Goodfellow, J.~Shlens, and C.~Szegedy, ``Explaining and harnessing
  adversarial examples,'' \emph{arXiv preprint arXiv:1412.6572}, 2014.

\bibitem{madry2017towards}
A.~Madry, A.~Makelov, L.~Schmidt, D.~Tsipras, and A.~Vladu, ``Towards deep
  learning models resistant to adversarial attacks,'' \emph{arXiv preprint
  arXiv:1706.06083}, 2017.

\bibitem{dziugaite2016study}
G.~K. Dziugaite, Z.~Ghahramani, and D.~M. Roy, ``A study of the effect of jpg
  compression on adversarial images,'' \emph{arXiv preprint arXiv:1608.00853},
  2016.

\bibitem{xie2017adversarial}
C.~Xie, J.~Wang, Z.~Zhang, Y.~Zhou, L.~Xie, and A.~Yuille, ``Adversarial
  examples for semantic segmentation and object detection,'' in
  \emph{Proceedings of the IEEE International Conference on Computer Vision},
  2017, pp. 1369--1378.

\bibitem{xie2017mitigating}
C.~Xie, J.~Wang, Z.~Zhang, Z.~Ren, and A.~Yuille, ``Mitigating adversarial
  effects through randomization,'' \emph{arXiv preprint arXiv:1711.01991},
  2017.

\bibitem{he2019parametric}
Z.~He, A.~S. Rakin, and D.~Fan, ``Parametric noise injection: Trainable
  randomness to improve deep neural network robustness against adversarial
  attack,'' in \emph{Proceedings of the IEEE Conference on Computer Vision and
  Pattern Recognition}, 2019, pp. 588--597.

\bibitem{MIT_survey}
N.~Akhtar and A.~Mian, ``Threat of adversarial attacks on deep learning in
  computer vision: A survey,'' \emph{IEEE Access}, vol.~6, pp.
  14\,410--14\,430, 2018.

\bibitem{yuan2019adversarial}
X.~Yuan, P.~He, Q.~Zhu, and X.~Li, ``Adversarial examples: Attacks and defenses
  for deep learning,'' \emph{IEEE transactions on neural networks and learning
  systems}, vol.~30, no.~9, pp. 2805--2824, 2019.

\bibitem{finn}
Y.~Umuroglu, N.~J. Fraser, G.~Gambardella, M.~Blott, P.~Leong, M.~Jahre, and
  K.~Vissers, ``Finn: A framework for fast, scalable binarized neural network
  inference,'' in \emph{Proceedings of the 2017 ACM/SIGDA International
  Symposium on Field-Programmable Gate Arrays}, 2017, pp. 65--74.

\bibitem{xcel_ram}
A.~Agrawal, A.~Jaiswal, D.~Roy, B.~Han, G.~Srinivasan, A.~Ankit, and K.~Roy,
  ``Xcel-ram: Accelerating binary neural networks in high-throughput sram
  compute arrays,'' \emph{IEEE Transactions on Circuits and Systems I: Regular
  Papers}, vol.~66, no.~8, pp. 3064--3076, 2019.

\bibitem{resparc}
A.~Ankit, A.~Sengupta, P.~Panda, and K.~Roy, ``Resparc: A reconfigurable and
  energy-efficient architecture with memristive crossbars for deep spiking
  neural networks,'' in \emph{Proceedings of the 54th Annual Design Automation
  Conference 2017}, 2017, pp. 1--6.

\bibitem{mehrotra1994fault}
K.~Mehrotra, C.~K. Mohan, S.~Ranka, and C.-t. Chiu, ``Fault tolerance of neural
  networks,'' SYRACUSE UNIV NY SCHOOL OF COMPUTER AND INFORMATION SCIENCE,
  Tech. Rep., 1994.

\bibitem{moons2016energy}
B.~Moons, B.~De~Brabandere, L.~Van~Gool, and M.~Verhelst, ``Energy-efficient
  convnets through approximate computing,'' in \emph{2016 IEEE Winter
  Conference on Applications of Computer Vision (WACV)}, 2016, pp. 1--8.

\bibitem{chakraborty2020constructing}
I.~Chakraborty, D.~Roy, I.~Garg, A.~Ankit, and K.~Roy, ``Constructing
  energy-efficient mixed-precision neural networks through principal component
  analysis for edge intelligence,'' \emph{Nature Machine Intelligence}, vol.~2,
  no.~1, pp. 43--55, 2020.

\bibitem{panda2020invited}
P.~Panda and K.~Roy, ``Invited talk: Re-engineering computing with
  neuro-inspired learning: Devices, circuits, and systems,'' in \emph{2020 33rd
  International Conference on VLSI Design and 2020 19th International
  Conference on Embedded Systems (VLSID)}, 2020, pp. 1--18.

\bibitem{venkataramani2015approximate}
S.~Venkataramani, S.~T. Chakradhar, K.~Roy, and A.~Raghunathan, ``Approximate
  computing and the quest for computing efficiency,'' in \emph{2015 52nd
  ACM/EDAC/IEEE Design Automation Conference (DAC)}, 2015, pp. 1--6.

\bibitem{bhattacharjee2020rethinking}
A.~Bhattacharjee and P.~Panda, ``Rethinking non-idealities in memristive
  crossbars for adversarial robustness in neural networks,'' \emph{arXiv
  preprint arXiv:2008.11298}, 2020.

\bibitem{quanos}
P.~Panda, ``Quanos-adversarial noise sensitivity driven hybrid quantization of
  neural networks,'' \emph{arXiv preprint arXiv:2004.11233}, 2020.

\bibitem{panda2019discretization}
P.~Panda, I.~Chakraborty, and K.~Roy, ``Discretization based solutions for
  secure machine learning against adversarial attacks,'' \emph{IEEE Access},
  vol.~7, pp. 70\,157--70\,168, 2019.

\bibitem{hw_trojans}
J.~Clements and Y.~Lao, ``Hardware trojan attacks on neural networks,''
  \emph{arXiv preprint arXiv:1806.05768}, 2018.

\bibitem{memory_attacks1}
Y.~Zhao, X.~Hu, S.~Li, J.~Ye, L.~Deng, Y.~Ji, J.~Xu, D.~Wu, and Y.~Xie,
  ``Memory trojan attack on neural network accelerators,'' in \emph{2019
  Design, Automation \& Test in Europe Conference \& Exhibition (DATE)}, 2019,
  pp. 1415--1420.

\bibitem{iia}
T.~A. Odetola, H.~R. Mohammed, and S.~R. Hasan, ``A stealthy hardware trojan
  exploiting the architectural vulnerability of deep learning architectures:
  Input interception attack (iia),'' \emph{arXiv preprint arXiv:1911.00783},
  2019.

\bibitem{dq}
J.~Lin, C.~Gan, and S.~Han, ``Defensive quantization: When efficiency meets
  robustness,'' \emph{arXiv preprint arXiv:1904.08444}, 2019.

\bibitem{vortex}
B.~Liu, H.~Li, Y.~Chen, X.~Li, Q.~Wu, and T.~Huang, ``Vortex: variation-aware
  training for memristor x-bar,'' in \emph{Proceedings of the 52nd Annual
  Design Automation Conference}, 2015, pp. 1--6.

\bibitem{domain_adv_training}
Y.~Ganin, E.~Ustinova, H.~Ajakan, P.~Germain, H.~Larochelle, F.~Laviolette,
  M.~Marchand, and V.~Lempitsky, ``Domain-adversarial training of neural
  networks,'' \emph{The Journal of Machine Learning Research}, vol.~17, no.~1,
  pp. 2096--2030, 2016.

\bibitem{tramer2017ensemble}
F.~Tram{\`e}r, A.~Kurakin, N.~Papernot, I.~Goodfellow, D.~Boneh, and
  P.~McDaniel, ``Ensemble adversarial training: Attacks and defenses,''
  \emph{arXiv preprint arXiv:1705.07204}, 2017.

\bibitem{athalye2018obfuscated}
A.~Athalye, N.~Carlini, and D.~Wagner, ``Obfuscated gradients give a false
  sense of security: Circumventing defenses to adversarial examples,''
  \emph{arXiv preprint arXiv:1802.00420}, 2018.

\bibitem{cisse2017parseval}
M.~Cisse, P.~Bojanowski, E.~Grave, Y.~Dauphin, and N.~Usunier, ``Parseval
  networks: Improving robustness to adversarial examples,'' \emph{arXiv
  preprint arXiv:1704.08847}, 2017.

\bibitem{ross2017improving}
A.~S. Ross and F.~Doshi-Velez, ``Improving the adversarial robustness and
  interpretability of deep neural networks by regularizing their input
  gradients,'' \emph{arXiv preprint arXiv:1711.09404}, 2017.

\bibitem{xie2019feature}
C.~Xie, Y.~Wu, L.~v.~d. Maaten, A.~L. Yuille, and K.~He, ``Feature denoising
  for improving adversarial robustness,'' in \emph{Proceedings of the IEEE
  Conference on Computer Vision and Pattern Recognition}, 2019, pp. 501--509.

\bibitem{chang2011priority}
I.~J. Chang, D.~Mohapatra, and K.~Roy, ``A priority-based 6t/8t hybrid sram
  architecture for aggressive voltage scaling in video applications,''
  \emph{IEEE transactions on circuits and systems for video technology},
  vol.~21, no.~2, pp. 101--112, 2011.

\bibitem{bortolotti2014approximate}
D.~Bortolotti, H.~Mamaghanian, A.~Bartolini, M.~Ashouei, J.~Stuijt, D.~Atienza,
  P.~Vandergheynst, and L.~Benini, ``Approximate compressed sensing: ultra-low
  power biosignal processing via aggressive voltage scaling on a hybrid memory
  multi-core processor,'' in \emph{2014 IEEE/ACM International Symposium on Low
  Power Electronics and Design (ISLPED)}, 2014, pp. 45--50.

\bibitem{hybridmem}
G.~Srinivasan, P.~Wijesinghe, S.~S. Sarwar, A.~Jaiswal, and K.~Roy,
  ``Significance driven hybrid 8t-6t sram for energy-efficient synaptic storage
  in artificial neural networks,'' in \emph{2016 Design, Automation \& Test in
  Europe Conference \& Exhibition (DATE)}, 2016, pp. 151--156.

\bibitem{liu2017neural}
Y.~Liu, Y.~Xie, and A.~Srivastava, ``Neural trojans,'' in \emph{2017 IEEE
  International Conference on Computer Design (ICCD)}, 2017, pp. 45--48.

\bibitem{kim2014flipping}
Y.~Kim, R.~Daly, J.~Kim, C.~Fallin, J.~H. Lee, D.~Lee, C.~Wilkerson, K.~Lai,
  and O.~Mutlu, ``Flipping bits in memory without accessing them: An
  experimental study of dram disturbance errors,'' \emph{ACM SIGARCH Computer
  Architecture News}, vol.~42, no.~3, pp. 361--372, 2014.

\bibitem{rakin2019bit}
A.~S. Rakin, Z.~He, and D.~Fan, ``Bit-flip attack: Crushing neural network with
  progressive bit search,'' in \emph{Proceedings of the IEEE International
  Conference on Computer Vision}, 2019, pp. 1211--1220.

\bibitem{chang20088t}
L.~Chang, R.~K. Montoye, Y.~Nakamura, K.~A. Batson, R.~J. Eickemeyer, R.~H.
  Dennard, W.~Haensch, and D.~Jamsek, ``An 8t-sram for variability tolerance
  and low-voltage operation in high-performance caches,'' \emph{IEEE Journal of
  Solid-State Circuits}, vol.~43, no.~4, pp. 956--963, 2008.

\bibitem{ozturk2008ilp}
O.~Ozturk and M.~Kandemir, ``Ilp-based energy minimization techniques for
  banked memories,'' \emph{ACM Transactions on Design Automation of Electronic
  Systems (TODAES)}, vol.~13, no.~3, pp. 1--40, 2008.

\bibitem{koc2006minimizing}
H.~Koc, O.~Ozturk, M.~Kandemir, S.~H.~K. Narayanan, and E.~Ercanli,
  ``Minimizing energy consumption of banked memories using data
  recomputation,'' in \emph{Proceedings of the 2006 international symposium on
  Low power electronics and design}, 2006, pp. 358--362.

\bibitem{whitley1994genetic}
D.~Whitley, ``A genetic algorithm tutorial,'' \emph{Statistics and computing},
  vol.~4, no.~2, pp. 65--85, 1994.

\bibitem{kennedy1995particle}
J.~Kennedy and R.~Eberhart, ``Particle swarm optimization,'' in
  \emph{Proceedings of ICNN'95-International Conference on Neural Networks},
  vol.~4, 1995, pp. 1942--1948.

\end{thebibliography}



%








\end{document}